\documentclass[journal]{IEEEtran}
\usepackage{latexsym,,amssymb,amsmath,graphicx,epsf,cite,bbm,float}
\usepackage{ifpdf}
\usepackage{epstopdf,mathtools}

\usepackage{algorithm,algorithmic}
\usepackage{amsmath,amssymb,bm}
\usepackage{amsfonts,dsfont,color,bbm,subcaption}

\usepackage{mathtools}

\def\ninept{\def\baselinestretch{1}}
\ninept

\newcommand{\abs}[1]{|#1|}

\DeclareMathOperator*{\argmin}{arg\,min}

\usepackage{hyperref,amsthm}

\newtheorem{theorem}{Theorem}

\newtheorem{lemma}[]{Lemma}

\newtheorem{proposition}[]{Proposition}

\newtheorem{remark}[]{Remark}

\newtheorem{definition}[]{Definition}

\newtheorem{assumption}[]{Assumption}

\begin{document}

\title{A Quadratic Time Locally Optimal Algorithm for NP-hard Equal Cardinality Partition Optimization} 
\author{\IEEEauthorblockN{Kaan Gokcesu}, \IEEEauthorblockN{Hakan Gokcesu} }
\maketitle

\begin{abstract}
	We study the optimization version of the equal cardinality set partition problem (where the absolute difference between the equal sized partitions' sums are minimized). While this problem is NP-hard and requires exponential complexity to solve in general, we have formulated a weaker version of this NP-hard problem, where the goal is to find a locally optimal solution. The local optimality considered in our work is under any swap between the opposing partitions' element pairs. To this end, we designed an algorithm which can produce such a locally optimal solution in $O(N^2)$ time and $O(N)$ space. Our approach does not require positive or integer inputs and works equally well under arbitrary input precisions. Thus, it is widely applicable in different problem scenarios.
\end{abstract}

\section{Introduction}\label{sec:intro}
\subsection{Preliminaries}
In the traditional set partition problem \cite{cook1971complexity,turing1937computable,levin1973universal,korf1998complete}, the goal is to decide whether a given input set $\mathcal{X}$ of positive integers can be partitioned into two disjoint complementary subsets $\mathcal{X}_1$ and $\mathcal{X}_2$ such that the sum of the elements in $\mathcal{X}_1$ equals to the sum of the elements in $\mathcal{X}_2$. A closely related problem to the set partitioning is the subset-sum problem \cite{kleinberg2006algorithm}, where the goal is to find a subset, sum of which equals a target value $T$ (which can be solved with a set partition solver by adding a dummy sample). In our paper, we investigate the equal cardinality partition problem, which is a variant where both partitions should have an equal number of items on top of having an equal sum. Specifically, we deal with the optimization version, which is to partition the set $\mathcal{X}$ into two disjoint subsets $\mathcal{X}_1$ and $\mathcal{X}_2$ of equal size such that the absolute difference between the sum of elements in $\mathcal{X}_1$ and the sum of elements in $\mathcal{X}_2$ is minimized. 

The traditional problem is one of Karp's $21$ NP-complete combinatorial problems \cite{karp1972reducibility} and one of Garey and Johnson’s $6$ NP-complete fundamental problems \cite{garey1979computers}. The equal cardinality variant is also NP-complete \cite{garey1979computers} and the optimization version is NP-hard \cite{cieliebak2002equal,korf2009multi}.
The traditional one has also been dubbed "the easiest hard problem" in literature because of its limited structure in comparison to other NP-complete problems \cite{hayes2002computing,mertens2006number}; because, despite its NP-completeness, there exist methods that can solve it optimally, approximately or locally in many instances (moreover, the set of viable solutions can be exponentially large). There exists pseudo-polynomial time algorithms, where the run time is polynomially dependent on the numeric values of the inputs in contrast to the length of the input \cite{garey1979computers}. There exists efficient locally optimal solvers for the traditional set partition \cite{gokcesu2021efficient}. All in all, it is a well-studied branch of combinatorial optimization with many important developments \cite{graham1966bounds,coffman1978application,dell1995optimal,korf1998complete,korf2009multi,moffitt2013search,schreiber2013improved,schreiber2014cached,schreiber2018optimal} and may imply similar possible improvements for harder NP-complete problems \cite{schreiber2018optimal}. 

\subsection{Notable Applications}
The set partitioning has a lot of related applications to problems in learning, optimization and decision theory, e.g., scheduling, allocation, classification and regression  \cite{neyshabouri2018asymptotically,cano2006combination,cano2007evolutionary,gokcesu2018adaptive,garcia2009enhancing,golbraikh2000predictive,gokcesu2020generalized,fan2005working,huberbook,cesabook,poor_book,kellerer2004knapsack,gokcesu2021generalized,mathews1896partition,dantzig2007number,coffman1984approximation,brams1996fair,biswas2018fair,gokcesu2021optimal,walsh2009really,merkle1978hiding,shamir1982polynomial,rivest1983cryptographic,gokcesu2021optimally,sarkar1987partitioning,dell2008heuristic,graham1979optimization,umang2013exact,lalla2016set,gokcesu2021regret}.

The knapsack problem \cite{kellerer2004knapsack} is one such closely related application, which governs the following problem: given a set of items with varying weights and values, determine the number of each said item in a collection so that the total weight is constrained by an upper bound and the total value is maximized. This problem often arises in resource allocation settings where we need to choose from a set of non-divisible tasks (or goods) under a time (or budget) constraint. The knapsack problem is a century-old problem \cite{mathews1896partition} and its name dates back to the early works of Dantzig \cite{dantzig2007number}. Its applications are numerous in real-world decision-making processes in a wide variety of fields, such as finding the least wasteful way to cut raw materials, selection of investments and portfolios, selection of assets for securitization, and generating keys for the Merkle–Hellman and other knapsack cryptosystems \cite{kellerer2004knapsack}.

The bin packing optimization problem \cite{coffman1984approximation} is another closely related one, where we need to pack certain items with varying sizes into a minimum number of bins with a fixed capacity. It has a myriad of applications from filling containers to loading trucks and creating file backups.

Another related problem is the fair division \cite{brams1996fair}, which is the problem of fairly dividing a set of resources among a certain number of recipients. It arises in many real-world scenarios such as division of inheritance, dissolution of partnerships, settlements in divorces, frequency allocation in band usage, traffic management of airports or harbors, satellite allocations and social choice theory \cite{brams1996fair,biswas2018fair}. 

Moreover, in some works, the partition problem is synonymous with parallel program scheduling \cite{dell2008heuristic}, where the goal is to assign various tasks to the parallel working machines while minimizing the total amount of time it takes to their completion \cite{garey1979computers,sarkar1987partitioning,dell2008heuristic,graham1979optimization}. A similar application is in berth allocation \cite{umang2013exact}, which aims to optimally allocate berth space for the incoming vessel in container terminals while minimizing the total time it takes to serve all the vessels \cite{umang2013exact,lalla2016set}. 

Fair team selection is a famous problem related to set partitions, where the goal is to create fair teams in regards to the total skill level of the players, i.e., minimizing the difference between the cumulative skill in different teams \cite{hayes2002computing}. Another famous example is for the manipulation of veto elections, where each elector has different veto weights and the candidate with the smallest total veto wins the election. Using set partitioning, a coalition can maximize their candidate's chances \cite{walsh2009really}.

\pagebreak
\subsection{Related Works} 
Given an input set $\mathcal{X}$ of size $N$, the most straightforward algorithm for the set partition problem is the brute force approach, which has $O(2^N)$ time complexity. The problem is NP-hard and its solution takes exponential time to find, however, there exist more efficient approaches. An example is the algorithm in \cite{horowitz1974computing}, which finds an optimal solution in $O(2^{N/2})$ time but also uses $O(2^{N/2})$ memory. While it has better run time than brute force, it also uses exponential memory. The algorithm in \cite{schroeppel1981t} addresses this issue and decreases the space complexity to $O(2^{N/4})$ while keeping the time complexity the same. Even though the problem is NP-hard, there are dynamic programming approaches as well \cite{garey1979computers,martello1990knapsack,korf2013optimally}, which can find an optimal solution in pseudo-polynomial time and space. Specifically, their complexities are polynomially dependent on the sum of inputs $S$. Note that, in such approaches the inputs are assumed to be positive and integer. Thus, their performance is highly dependent on the precision of the inputs. There also exists sub-optimal algorithms like the greedy algorithm and the Karmarkar-Karp set differencing algorithm, which run in $O(N\log N)$ time and $O(N)$ space \cite{graham1966bounds,karmarkar1982differencing,kellerer2004knapsack,korf2011hybrid}. The complete anytime algorithm in \cite{korf1998complete} can transform such efficient sub-optimal algorithms into optimal ones. It creates a binary tree from the sub-optimal algorithm selections (much like a modification of the brute force approach) to create anytime algorithms with linear memory usage. However, its worst-case time complexity is $O(2^N)$.
There are also algorithms which can produce locally optimal solutions for the set partition problem in $O(N\log N)$ time and $O(N)$ space \cite{gokcesu2021efficient}. Instead of a seemingly arbitrary sub-optimal heuristic, a locally optimal solver may prove to be more useful in many scenarios.

\subsection{Contributions and Organization}
Even though there exists optimal algorithms, they have exponential time (albeit better than brute force) and space complexities. There also exists an optimal solver with linear $O(N)$ memory, however, its time complexity is the same as brute force in the worst case. For efficient approaches, dynamic programming methods find an optimal one with pseudo-polynomial complexity. However, it has limited use for high precision or non-integer inputs. Although there are linearithmic time complexity algorithms, they generally produce sub-optimal solutions. While there exists a locally optimal algorithm for the set partitioning, it is not applicable to the equal cardinality variant. 
Improving the exponential complexity is unfruitful because of the NP-hardness of the problem but fast algorithms are always desired especially with the emergence of big data. To this end, much like \cite{gokcesu2021efficient}, we postulate that finding a 'locally' optimal solution to the equal cardinality set partition problem is nowhere near as hard as finding a 'globally' optimal one. 
In \autoref{sec:prob}, we mathematically formulate the problem definition. In \autoref{sec:method}, we design an algorithm and prove that its locally optimal solution takes $O(N^2)$ time and $O(N)$ space to find. In \autoref{sec:disc}, we provide some important discussions, show the extension to the traditional set partition problem and conclude with a few remarks.

\section{Locally Optimal Equal Cardinality Partition}\label{sec:prob}
In this section, we formally define the equal cardinality set partition problem. As an input, we have the set of numbers
	\begin{align}
	\mathcal{X}=&\{x_1,x_2,\ldots,x_N\},\\
	=&\{x_n\}_{n=1}^N,
	\end{align}	
where $N$ is an even number.

We partition $\mathcal{X}$ into two equally sized subsets $\mathcal{X}_1$ and $\mathcal{X}_2$ such that they are disjoint and their union is $\mathcal{X}$, i.e.,
	\begin{align}
	\mathcal{X}_1\cap\mathcal{X}_2&=\emptyset,\\
	\mathcal{X}_1\cup\mathcal{X}_2&=\mathcal{X},\\
	\abs{\mathcal{X}_1}=\abs{\mathcal{X}_2}&=\frac{N}{2},
	\end{align}	
where $\abs{\cdot}$ is the cardinality.
The goal is to create the sets $\mathcal{X}_1$ and $\mathcal{X}_2$ such that their individual sums are as close as possible to each other, where the set sums are given by
\begin{align}
S_1=\sum_{x\in\mathcal{X}_1}x,\\
S_2=\sum_{x\in\mathcal{X}_2}x,
\end{align}
Hence, $S_1$ and $S_2$ are each other's complements, i.e., $S_1=S-S_2$ and $S_2=S-S_1$, where
\begin{align}
	S=\sum_{x\in\mathcal{X}}x.
\end{align}
We consider the formulation where we want to minimize the absolute difference between the sums $S_1$ and $S_2$, i.e.,
\begin{align}
\min_{\mathcal{X}_1,\mathcal{X}_2} \left(\abs{S_1-S_2}\right).
\end{align}
However, we point out that the formulation of minimizing the maximum sum, i.e.,
\begin{align}
\min_{\mathcal{X}_1,\mathcal{X}_2} \left(\max(S_1,S_2)\right),
\end{align}
or the formulation of maximizing the minimum sum, i.e.,
\begin{align}
\max_{\mathcal{X}_1,\mathcal{X}_2} \left(\min(S_1,S_2)\right),
\end{align}
are all equivalent. Since we are only dealing with two sets and their corresponding sums, in general, all such formulations become equivalent to each other \cite{gokcesu2021efficient}.

This problem is NP-hard \cite{karp1972reducibility} and impossible to solve with an efficient method. To this end, instead of this NP-hard problem, we consider a 'weaker' version of the equal cardinality set partition problem. Instead of a global optimal solution, we are after a 'locally' optimal one, which is defined as follows:

\begin{definition}
	A partition $(\mathcal{X}_1,\mathcal{X}_2)$ of $\mathcal{X}$ is locally optimal if there is no single swap of element pairs (belonging to different sets) that can decrease the absolute difference between the set sums $S_1$ and $S_2$, i.e.,
	\begin{align}
		\abs{(S_1-x_1+x_2)-(S_2+x_1-x_2)}\geq&\abs{S_1-S_2},
	\end{align}
	for all $x_1\in\mathcal{X}_1$ and $x_2\in\mathcal{X}_2$.
\end{definition}

Note that this local optimality definition is fundamentally different from \cite{gokcesu2021efficient}. Theirs was under any single element transfer between the partitions, which is not meaningful for equal cardinality case.

\section{Finding A Locally Optimal Solution to Equal Cardinality Set Partition in Quadratic Time}\label{sec:method}
\subsection{Iterative Algorithm}\label{sec:alg}
Before we propose the algorithm, we make some initial assumptions.
\begin{assumption}
	Let the set $\mathcal{X}$ be composed of only positive elements, i.e.,
	\begin{align*}
		x>0, &&\forall x\in \mathcal{X}.
	\end{align*}
\end{assumption}
\begin{assumption}
	Let the set $\mathcal{X}=\{x_n\}_{n=1}^N$ be in ascending order, i.e.,
	\begin{align*}
	x_n\leq x_{n+1}, &&\forall n\in\{1,2,\ldots,N-1\}.
	\end{align*}
\end{assumption}
Thus, our input $\mathcal{X}$ is a positive ordered set. However, we point out that there is no requirement for the elements to be integers, i.e., they can be reals with arbitrary precisions. 

Given the input set $\mathcal{X}$, the algorithm works as follows:
\begin{enumerate}
	\item We start with an arbitrary partition of $\mathcal{X}$: $\mathcal{X}_1$ and $\mathcal{X}_2$ with the initial sums $S_1$ and $S_2$ respectively.\label{step:start}
	\item Set $n=1$\label{step:n}
	\item Check if swapping $x_n$ with any element $x_{n'}$ ($n'<n$) from the opposing set strictly decreases the absolute sum difference $\abs{S_1-S_2}$. \label{step:xn}
	\item If yes, continue; \\
	else if $n=N$, STOP;
	\\ or else, set $n\leftarrow n+1$ and return to Step \ref{step:xn}.
	\item Swap $x_n$ with the element in the other set that most decreases the absolute sum difference. For example, if $x_n\in\mathcal{X}_1$, swap $x_n$ with $x_{n_*}$, where $$n_*=\argmin_{n'<n:x_{n'}\in\mathcal{X}_2}|S_1-S_2-2x_n+2x_{n'}|,$$ and make the necessary updates
	\begin{align}
		\mathcal{X}_1&\leftarrow (\mathcal{X}_1\setminus\{x_n\})\cup\{x_{n_*}\},\\ \mathcal{X}_2&\leftarrow (\mathcal{X}_2\setminus\{x_{n_*}\})\cup\{x_{n}\},\\ S_1&\leftarrow S_1-x_n+x_{n_*},\\ 
		S_2&\leftarrow S_2-x_{n_*}+x_n.
	\end{align} 
	\item If the sign of $S_1-S_2$ remains unchanged,
	\subitem if $n=N$, STOP;
	\subitem or else, set $n\leftarrow n+1$ and return to Step \ref{step:xn};
	\\or else, return to Step \ref{step:n}.\label{step:it}
\end{enumerate}

\begin{remark}
	Heuristics can be used in Step \ref{step:start} to speed up the algorithm. Initial state is inconsequential in the worst-case.
\end{remark}

\begin{proposition}
	The algorithm definitely terminates.
	\begin{proof}
		The algorithm terminates, i.e., stops, only when $n=N$ in the algorithm. When $n<N$, it either returns to Step \ref{step:xn}; or it returns to Step \ref{step:n} if the sign of $S_1-S_2$ changes after a swap. We know that with each swap the absolute difference $\abs{S_1-S_2}$ strictly decreases because of Step \ref{step:xn}. Since there are a finite number of elements, hence, finite possible swaps; the sign of $S_1-S_2$ cannot keep changing indefinitely. Thus, the return to Step \ref{step:n} happens a finite amount of times. Together with the fact that whenever we return to Step \ref{step:xn}, we have already increased $n$ by $1$; the algorithm reaches $n=N$ after a finite number of iterations and definitely terminates.
	\end{proof}
\end{proposition}
 
\subsection{Local Optimality}
In this section, we prove the local optimality of our algorithm, where we start with a few useful results.

\begin{proposition}
	The absolute sum difference can only be decreased by swapping two elements, where the larger one belongs to the set with the larger sum.
	\begin{proof}
		Without loss of generality let $S_1-S_2>0$. For the absolute sum difference $\abs{S_1-S_2}$ to decrease the swapped elements $x_1$ and $x_2$ should have the inequality $x_1>x_2$. Otherwise, $S_1-S_2$ will get larger and so is $\abs{S_1-S_2}$.
	\end{proof}
\end{proposition}

\begin{proposition}\label{thm:xn1xn2}
	If swapping elements in different sets $x_{n_1}$ and $x_{n_2}$ for some $n_1>n_2$ does not decrease the absolute difference of set sums $\abs{S_1-S_2}$; neither does swapping $x_{n_1'}$ (for some $n_1'\geq n_1$) with $x_{n_2'}$ (for some $n_2'\leq n_2$).
	\begin{proof}
		The proof is straightforward from the fact that $\{x_n\}_{n=1}^N$ is in ascending order, since the difference ${x_{n_1}-x_{n_2}}$ is less than or equal to ${x_{n_1'}-x_{n_2'}}$ for any $n_1'\geq n_1$ and $n_2'\leq n_2$.
	\end{proof}
\end{proposition}

\begin{proposition}\label{thm:SoldSnew}
	Without loss of generality let $S_1^{old}-S_2^{old}=\triangle^{old}>0$. If swapping elements in different sets $x_{1}\in\mathcal{X}_1$ and $x_{2}\in\mathcal{X}_2$ ($x_1>x_2$) did not decrease the absolute difference of set sums; neither does swapping them when $S_1^{new}-S_2^{new}=\triangle_{new}>0$, where $\triangle_{new}\leq\triangle_{old}$.
	\begin{proof}
		Since the swap did not decrease the absolute difference of set sums, we have $x_1-x_2\geq \triangle_{old}$, which also implies $x_1-x_2\geq \triangle_{new}$ and concludes the proof.
	\end{proof}
\end{proposition}

\begin{proposition}\label{thm:xn2}
	Without loss of generality let $S_1-S_2>0$. Let $x_{n_1}\in\mathcal{X}_1$, $x_{n_2}\in\mathcal{X}_2$ (such that $n_1>n_2$) be a pair of elements whose swap maximally decreases the absolute difference of set sums over all possible swaps including $x_{n_1}$. After the swap, there are no elements $x_{n_3}\in\mathcal{X}_2$ ($n_2>n_3$), whose swap with $x_{n_2}$ strictly decreases the absolute difference of set sums.
	\begin{proof}
		The proof comes from the fact that the pair $x_{n_1}$ and $x_{n_2}$ is a minimizer swap over the swaps including $x_{n_1}$. If there indeed were a feasible $x_{n_3}$, the minimizer swap would simply be between $x_{n_1}$ and $x_{n_3}$.
	\end{proof}
\end{proposition}

\begin{lemma}
	At the stop, we have a local optimal solution.
	\begin{proof}
		At the algorithm's termination, we make a final traverse from $n=1$ to $n=N$ all the while never returning to Step \ref{step:n} from Step \ref{step:it}. In this traverse, $n$ increases by $1$ if swapping $x_n$ with any $x_{n'}$ ($n'<n$) is unfruitful or we make a swap and the sign of $S_1-S_2$ remains unchanged. Without loss of generality let $S_1-S_2>0$. If there are no swaps in the traverse, that means there are no pairs $x_{n_1}\in\mathcal{X}_1$ and $x_{n_2}\in\mathcal{X}_2$ ($n_1>n_2$) whose swap decreases the absolute difference of set sums, which implies the local optimality. Suppose there is a swap between $x_{n_1}\in\mathcal{X}_1$ and $x_{n_2}\in\mathcal{X}_2$ ($n_1>n_2$). After the swap, from \autoref{thm:SoldSnew}, we have that $x_n\in\mathcal{X}_1$ for $n<n_1$ and $n\neq n_2$ still has no viable swaps. Moreover, from \autoref{thm:xn2}, we have that $x_{n_2}$ has no viable swap. Thus, $x_{n}\in\mathcal{X}_1$ for $n<n_1$ has no viable swap. Henceforth, in the final traverse, the swaps eliminate any remaining viable swaps while not affecting the smaller elements. At $n=N$ we will end up with a local optimal solution.
	\end{proof}
\end{lemma}

\subsection{Complexity Analysis}

Since our algorithm only keeps track of which elements are in which partition, it uses $O(N)$ space. 

\begin{lemma}\label{thm:X2}
	In the algorithm, whenever a swap happens between $x_{n_1}\in\mathcal{X}_1$ and $x_{n_2}\in\mathcal{X}_2$ (without loss of generality let $S_1-S_2>0$ and $n_1>n_2$), we have before the swap: $x_{n_1-1}\in\mathcal{X}_2$ and $\max_{n<{n_1}:x_n\in\mathcal{X}_1}n< {n_2}$.
	\begin{proof}
		The algorithm checks $x_{n_1}\in\mathcal{X}_1$ only if $x_{n<n_1}\in\mathcal{X}_1$ has no viable swaps. Together with \autoref{thm:xn1xn2} concludes the proof.
	\end{proof}
\end{lemma}

\begin{lemma}\label{thm:cons}
	Any swap in the algorithm can be expressed as a series of swaps between consecutive elements.
	\begin{proof}
		From \autoref{thm:X2}, we have that whenever there is a swap between $x_{n_1}\in\mathcal{X}_1$ and $x_{n_2}\in\mathcal{X}_2$ (without loss of generality let $S_1-S_2>0$ and $n_1>n_2$), all the elements between them are in $\mathcal{X}_2$, i.e., $x_{n_1>n>n_2}\in\mathcal{X}_2$. Thus, this swap is equivalent to swapping $x_{n_1-i}$ and $x_{n_1-i-1}$ consecutively for $i\in\{0,1,\ldots,n_1-n_2-1\}$, which concludes the proof.
	\end{proof}
\end{lemma}

\begin{lemma}\label{thm:decAbs}
	In the algorithm, whenever a sign change happens on $S_1-S_2$ after a swap between $x_{n_1}\in\mathcal{X}_1$ and $x_{n_2}\in\mathcal{X}_2$ (without loss of generality let $S_1-S_2>0$ and $n_1>n_2$); we have $\abs{S_1-S_2-2x_{n_1}+2x_{n_2}}\leq \abs{x_{n_2+1}-x_{n_2}}$.
	\begin{proof}
		The proof comes from \autoref{thm:cons}. Since a series of swaps between consecutive elements results in the sign change. The last swap in the series results in the sign change, which concludes the proof.
	\end{proof}
\end{lemma}

\begin{lemma}\label{thm:Nite}
	In the algorithm, each traverse over $n$ (i.e., from $n=1$ to again $n=1$) takes $O(N)$ time.
	\begin{proof}
	Without loss of generality let $S_1-S_2>0$ in our traverse (note that whenever the sign changes we begin a new traverse). Suppose we are at some time iteration $n_1$ and we have $\max_{n<{n_1}:x_n\in\mathcal{X}_1}n=n_1'<n_1$. This implies $x_{n'\leq n_1'}\in\mathcal{X}_1$ has no viable swaps. From \autoref{thm:X2}, we know that $x_{n_1}$ can only be swapped with $x_{n_1'<n<n_1}$ (all of which are in $\mathcal{X}_2$). We start by checking the swap with $x_{n=n_1'+1}$ and do a simple linear search to find the best swap, which takes $O(n_1-n_1')$ time. Since we iteratively do this for all elements in $\mathcal{X}_1$ in a single traverse, it takes at most $O(N)$ time, which concludes the proof.
	\end{proof}
\end{lemma}

\begin{lemma}\label{thm:Ntra}
	In the algorithm, we have at most $O(N)$ traverses.
	\begin{proof}
		A traverse end after a swap which causes a sign change in $S_1-S_2$. From \autoref{thm:decAbs}, we have that whenever the sign changes, the resulting absolute difference between the set sums will be less than the difference between some consecutive elements in $\mathcal{X}$. Since with each swap, the absolute difference decreases, this consecutive swap can never happen again after it has happened once. Since there are $N-1$ possible consecutive element pairs, the number of traverse can never exceed $O(N)$, which concludes the proof.
	\end{proof}
\end{lemma}

\begin{theorem}
	Our algorithm takes $O(N^2)$ time.
	\begin{proof}
		The result comes from combining the results of \autoref{thm:Nite} and \autoref{thm:Ntra}.
	\end{proof}
\end{theorem}

\section{Discussions and Conclusion}\label{sec:disc}

Our algorithm does not use the fact that $x_n$ are positive, hence, it is redundant. However, since we are dealing with the equal cardinality set partition problem, note that, any arbitrary scaling or translation has no effect on the partitioning. Thus, any affine transform of $x_n$ in the form of $\alpha x_n+\beta$ for nonzero $\alpha$ produces the same partitioning. The invariance under arbitrary scalings is true for any possible set partition not just equal cardinality. However, note that, the translation invariance only exists for the partitions with equal cardinality in general, since the effect of $\beta$ will be $\frac{N\beta}{2}$ on both partitions.

If $\mathcal{X}$ is unordered, it takes $O(N\log N)$ time and $O(N)$ space to order it, which does not change our complexity since our algorithm runs in $O(N^2)$ time and $O(N)$ space. 

Note that our local optimality definition is fundamentally different from \cite{gokcesu2021efficient}, which studies general set partitions. The results in \cite{gokcesu2021efficient} are locally optimal under any element transfers between the partitions, which is not applicable in our problem, since the equal cardinality partition does not allow an element to transfer between partitions by itself. To this end, our solutions are locally optimal under any swap between elements.

We point out that our algorithm can be straightforwardly extended to the case where we search for a local optimal partitioning with possibly different cardinalities as in \cite{gokcesu2021efficient}. We first observe there is a simple reduction from the equal cardinality variant to the traditional set partitioning. Any algorithm that can solve the equal cardinality problem can be used for the traditional set partition problem by addition of some dummy zero elements. Hence, the equal cardinality variant is at least as hard as the traditional set partition problem. To apply our algorithm in the traditional set partition problem, we need to extend the input set $\mathcal{X}$ of size $N$ to a set of size $2N$, which includes all the elements in $\mathcal{X}$ and $N$ zeros. After running our algorithm on this set, we will end up with a partitioning that has a stronger local optimality than \cite{gokcesu2021efficient}. Since in this case, the solution will be not only optimal with regards to any single element transfers but also any swap between element pairs in different sets. Note that, the stronger optimality comes together with an increase in the complexity of the algorithm.

Moreover, it is straightforward to apply our algorithm to not just the equal cardinality setting but any predefined cardinalities. Since our algorithm works by swapping element between partitions, the initial cardinalities will be preserved.

In conclusion, we have studied the optimization version of the equal cardinality set partition problem (where the absolute difference between the partition sums are minimized and the sets have equal cardinality). While this problem is NP-hard and requires exponential complexity to solve in general, we have formulated a weaker version of this NP-hard problem, where the goal is to find a locally optimal solution. The local optimality considered in our work is under any single swap between the opposing partition elements. To this end, we designed an algorithm which can produce such a locally optimal solution in $O(N^2)$ time and $O(N)$ space. Our approach does not require positive or integer inputs. Our algorithms work equally well under arbitrary input precisions, hence, widely applicable in difference problem scenarios and settings.

\bibliographystyle{IEEEtran}
\bibliography{double_bib}

\end{document}